\documentclass[aps,pra,twocolumn,superscriptaddress,showpacs,floatfix]{revtex4}
\usepackage[dvipdfmx]{graphicx,color}
\begin{document}

\title{Loading of atoms into an optical trap with high initial phase-space density}
\author{Kosuke Shibata}
\email[Present address: 
Gakushuin University, Department of Physics, 1-5-1 Mejiro, Toshima-ku, Tokyo, Japan.
E-mail address: ]
{shibata@qo.phys.gakushuin.ac.jp}
\author{Shota Yonekawa}
\author{Satoshi Tojo}
\affiliation{Department of Physics, 
Chuo University, 1-13-27 Kasuga, Bunkyo, Tokyo 112-8551, Japan}
\date{\today}

\begin{abstract}
We report a method for loading cold atoms into an optical trap
with high initial phase-space density (PSD).
When the trap beam is overlapped with atoms in optical molasses
of optimized parameters including large cooling beam detuning 
compared with conventional detuning used for a magneto-optical trap (MOT), 
more than $3 \times 10^6$ rubidium atoms 
with an initial temperature less than 20 $\mu$K are loaded into a single beam trap.
The obtained maximum initial PSD is estimated to be $1.1 \times 10^{-3}$,
which is one or two orders of magnitude greater than that achieved with 
the conventional loading into an optical trap from atoms in a MOT.
The proposed method is promising 
for creating a quantum gas with a large number of atoms in a short evaporation time.

\end{abstract}

\maketitle
\section{Introduction}
An optical trap,
in which a detuned light beam confines an atom due to AC Stark potential, 
is one of the most useful tools in cold-atom experiments \cite{Grimm200095}.
It can hold atoms regardless of their spin, whereas a magnetic trap can hold only atoms in low-field-seeking states.
Owing to this advantage,
the optical trap has enabled 
the study of collisional properties between different spin states \cite{PhysRevA.80.042704} or spinor Bose-Einstein condensates (BECs) 
\cite{RevModPhys.85.1191, PhysRevLett.81.1539, Stenger1998, PhysRevLett.92.140403}.
Another advantage of optical traps is the high flexibility for spatial shapes in the trap.
Trap potentials of various shapes, including box-shaped trap potentials \cite{PhysRevLett.110.200406}, can be created via suitably designed optical traps.
Optical traps also enable dynamical shape change using 
a lens on a moving stage \cite{PhysRevLett.88.020401},
an acousto-optic modulator (AOM) \cite{PhysRevA.93.043403},
and a focus-tunable lens \cite{1367-2630-16-9-093028}.
Moreover, evaporative cooling in an optical trap
requires a shorter time compared to that in a magnetic trap
owing to the tight confinement in optical traps.
The short evaporation time enables a high repetition rate in experiments
and relaxes the requirement for a high vacuum, which is mandatory for achieving a long trap time.
In addition, the evaporative cooling in optical traps is indispensable 
for creating quantum degenerate gases of atoms having no electric spins
such as alkaline-earth (like) species
(for example, \cite{PhysRevLett.91.040404, PhysRevLett.94.160401}).

The simple and most popular strategy to produce quantum degenerated gases
through an all-optical method (that is, without magnetic traps)
has been the initial loading of many atoms from a magneto-optical trap (MOT)
into a deep trap produced by a high-power beam. 
Although the atoms get heated during the loading 
and their initial temperature is higher than that in MOTs,
the trapped atoms can be cooled via successive evaporation.
If the initial number of atoms in the trap is sufficient,
the atoms reach quantum degeneracy with evaporative cooling. 
A BEC containing $10^{4}$ or $10^{5}$ atoms is typically produced
when a few million bosonic atoms are initially loaded into an optical trap
with an evaporative cooling time of several seconds
\cite{PhysRevLett.87.010404,Arnold20113288,JPSJ.81.084004}.
A larger quantum degenerate gas will be created in a shorter evaporation time
if the initial temperature can be decreased.
Kinoshita \textit{et al.} \cite{PhysRevA.71.011602} reported
the production of a BEC with $3.5 \times 10^5$ atoms through the evaporative cooling in an optical trap in 1 s with pre-cooling in an optical lattice before loading the optical trap.

We report a simple and effective method for loading many 
well-cooled atoms in an optical trap.
The loading is performed by overlapping the beam with atoms in optical molasses.
More than three million atoms are loaded into a single-beam optical trap 
with the depth of 280 $\mu$K.
The initial temperature of loaded atoms is found to be 15 $\mu$K,
while the temperature in the conventional loading method 
is accompanied by heating,
after which the temperature settles to $\sim 1/10$ of the trap depth
by plain evaporation with loss of a fraction of atoms \cite{Arnold20113288}.
Owing to the low temperature compared with the trap depth,
the estimated initial PSD for the optimal loading exceeds $10^{-3}$,
which is one or two orders of magnitude greater than that obtained with the conventional loading method.
We attribute the increased PSD to 
the absence of a magnetic field gradient during the loading into the optical trap
and the detuning of the cooling beam much larger than that used for conventional MOT.

\section{Experimental Setup}\label{sec: setting}
We prepare cold $^{87}$Rb atoms in a vacuum glass cell
using a MOT in the independent six $\sigma^{+} \sigma^{-}$ beam configuration.
The cooling beam is detuned by $\Delta_\mathrm{M}$ = $-20$ MHz from 
the 5$^2$\textit{S}$_{1/2}$ $\textit{F}$ = 2 to 5$^2$\textit{P}$_{3/2}$ $\textit{F}'$ = 3 transition and the repump beam is resonant to the $\textit{F}$ = 1 to $\textit{F}'$ = 2 transition.
The cooling and repump beams are output from an optical fiber, 
expanded to a beam radius of 8 mm, 
and divided into six beams with equal power.
The powers of the cooling and repump beams output from the fiber are 150 mW and 2 mW, respectively.
The peak intensities at the cell (summed over all beams) 
are $I_\mathrm{M}$ = 140 mW/cm$^2$ and $I_\mathrm{R}$ = 2 mW/cm$^2$, respectively,
considering the power loss at the cell.
A pair of coils in the anti-Helmholtz configuration generates
a magnetic-field gradient of 11 G/cm (along the coil axis) for the MOT.
We typically collect $2 \times 10^8$ atoms in the MOT within 10 s.

After collecting atoms, we compress the MOT by using the following standard method.
We increase the field gradient to 25 G/cm in 10 ms and
decrease the repump beam intensity $I_\mathrm{R}$ to 10 $\mu$W/cm$^2$ in another 10 ms.
During the decrease of repump beam intensity, 
$\Delta_\mathrm{M}$ is swept to $-32$ MHz
by changing the RF frequency applied to 
an AOM in the cooling beam path.

We turn off the magnetic-field gradient within 1 ms after the compression
and start the polarization gradient cooling (PGC),
which consists of two parts.
In the first part, $\Delta_\mathrm{M}$ is swept from  $-32$ MHz to $-80$ MHz 
and $I_\mathrm{M}$ is decreased to 1/4 of the initial intensity in 10 ms
while $I_\mathrm{R}$ is kept constant.
The second part starts with a frequency jump of the laser locking point
from the $\textit{F}'$ = 3 peak to the $\textit{F}'$ = 2 - 3 crossover peak.
This jump shifts $\Delta_\mathrm{M}$ by $-133$ MHz and enables large detuning.
We use commercially available laser servo controller (Vescent Photonics Inc., D2-125),
which realizes the frequency jump with a sample-and-hold circuit.
The frequency jump is achieved within 600 $\mu$s, 
during which no significant diffusion of atoms is observed.
We note that the jump almost perfectly succeeds with optimization 
of the locking signal, the voltage determining the jumping frequency and the unlocking duration.
After the jump, the repump beam detuning is set to 
$\Delta_\mathrm{R} = -123$ MHz
to be led to the gray molasses cooling \cite {PhysRevLett.75.37, PhysRevA.53.R3734}. 
The repump beam power is increased to the maximum in this period.
At the same time $I_\mathrm{M}$ and $\Delta_\mathrm{M}$ are swept 
and their final values are 15 mW/cm$^2$ and $-240$ MHz, respectively. 
The cooling in this part lasts for 2 ms.
After the second part,
$1.8 \times 10^8$ atoms at less than 10 $\mu$K are typically produced
with cancellation of the residual magnetic field using three coil pairs
for generating conditions that allow effective PGC. 
The coil currents are finely adjusted 
to minimize the temperature in the molasses after the second part cooling.
We estimate the residual field to be less than 50 mG.

The optical trap beam is turned on after the atom cloud is cooled
with the cooling and repump beams kept on.
The trap beam is generated from a Ti:S laser and is red-detuned from the D$_2$ line by 2.3 THz.
The beam is passed through an AOM for intensity control and
delivered to the cell via an optical fiber.
The beam profile is elliptic 
with the vertical and horizontal beam waists being 23 $\mu$m and 69 $\mu$m, respectively.
The beam is focused in the molasses.
Further, the beam propagates in the horizontal plane 
and is linearly polarized along the horizontal axis.
The beam power is set at the maximum value of 196 mW,
unless otherwise indicated.
The maximum potential depth is calculated to be $k_\mathrm{B} \times$ 280 $\mu$K.

The molasses and the optical trap beam are overlapped for 40 ms,
unless otherwise indicated.
The loading of atoms is completed in this period, as described later.
After the loading, we shut off the cooling and repumping beams
and hold atoms in the optical trap for 25 ms, 
during which the atoms in the molasses fall off and are separated from the optical trap.
During the holding, the atoms are irradiated by a weak pumping beam, and almost all atoms are pumped to the $\textit{F}$ = 1 state.
We measure the atom number and temperature 
of the atoms in the optical trap as well as those in the MOT and the molasses
through absorption imaging after the time of flight. 

\section{Dependence of loading on molasses parameters}
We investigate the dependence of the loading on molasses beam parameters
such as the intensity and detuning of the cooling and repump beams.
The molasses parameters during the second PGC part and the loading
are changed when taking the following data.

\begin{figure}[t]
 \centering
  \includegraphics[width=7cm]{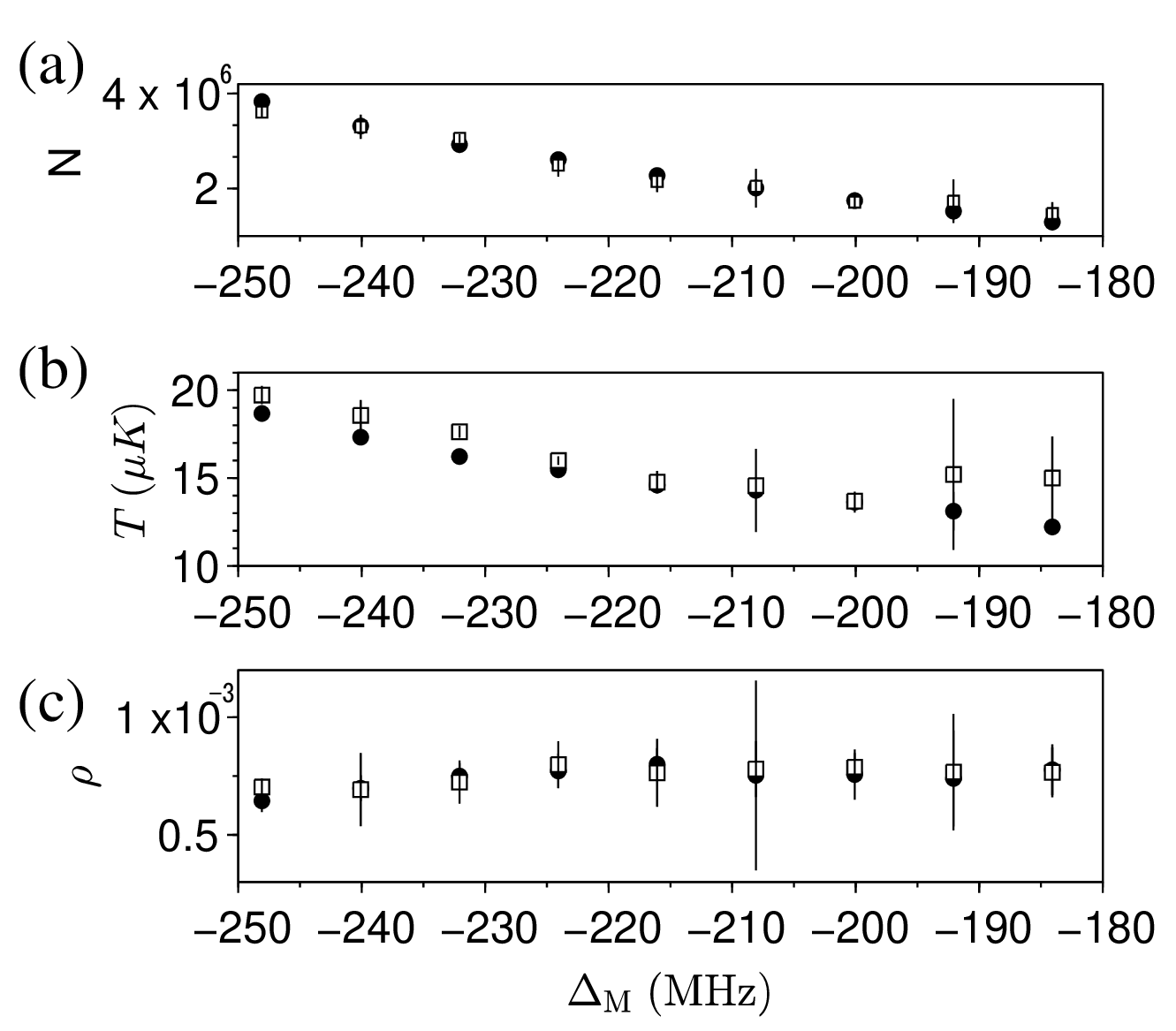}
 \caption{Cooling-beam-frequency dependence of 
the number (a) , temperature (b) 
and phase space density (c) of trapped atoms.
The circles and squares are for 
$I_\mathrm{M}$ = 15 mW/cm$^2$ and 42 mW/cm$^2$,
respectively.}
 \label{fig:cooling-dependence}
\end{figure}

We first focus on the dependence
of the atom number and temperature on the cooling beam detuning $\Delta_\mathrm{M}$.
The repump beam intensity and detuning are fixed during this measurement
to $I_\mathrm{R}$ = 10 $\mu$W/cm$^2$ and $\Delta_\mathrm{R}$ = $0$ MHz, respectively,
which are optimal values for the loading, as shown later.
Figure \ref{fig:cooling-dependence} (a) shows
the atom number increases with increasing $\vert \Delta_\mathrm{M} \vert$ in the measured detuning range.
The data with larger detuning is not taken owing to technical limitations.
The atom numbers are almost same for 
$I_\mathrm{M}$ = 15 mW/cm$^2$ and 42 mW/cm$^2$.

The temperature increases for large detuning  (see Fig. \ref{fig:cooling-dependence} (b)).
The temperature for $I_\mathrm{M}$ = 15 mW/cm$^2$ is 
slightly lower than that for $I_\mathrm{M}$ = 42 mW/cm$^2$ when compared for the same $\Delta_\mathrm{M}$.
This behavior of the temperature is reasonable 
because the laser frequency becomes close to the $\textit{F}$ = 2 - $\textit{F}'$ = 2 resonance, which is separated by 266 MHz from 
the $\textit{F}$ = 2 - $\textit{F}'$ = 3 transition frequency.
In addition, the AC stark effect in the optical trap shifts the resonance frequency
by approximately $+$ 10 MHz for the deepest trap;
consequently, the laser frequency becomes even closer 
to the shifted $\textit{F}$ = 2 - $\textit{F}'$ = 2 resonance. 
Although the larger detuning is advantageous to maximize the atom number,
the large detuning causes temperature increase. 
The PSD is almost constant when $\vert \Delta_\mathrm{M} \vert$ is smaller than approximately  230 MHz
and it drops for the larger detuning as shown in Fig. \ref{fig:cooling-dependence} (c).
The following experiments are performed 
with $I_\mathrm{M}$ = 28 mW/cm$^2$ and $\Delta_\mathrm{M}$ = $-232$ MHz
with experimental optimization,
which gives a high PSD and large atom number as well.

It should be noted that 
the low temperature of the atoms in the optical trap demonstrated here is quite striking
compared with the loading from MOT.
The conventional loading method is usually accompanied with the heating of atoms 
(an initial temperature $T \sim U/4$ has been reported \cite{PhysRevA.62.013406}),
after which the temperature settles to $\sim U/10$ due to evaporation. 
The temperature here is well less than 1/10 of the trap depth,
which is kept constant to $k_\mathrm{B} \times$ 280 $\mu$K
for these data.
Considering the deformation of the trap due to 
the scattering force \cite{PhysRevLett.57.314}, 
the net potential depth is $U = k_\mathrm{B} \times$ 230 $\mu$K
and $T$ is still less than $U/10$.
This result indicates that the atoms in the optical trap
are cooled by the PGC during the loading.

\begin{figure}[t]
 \centering
  \includegraphics[width=7cm]{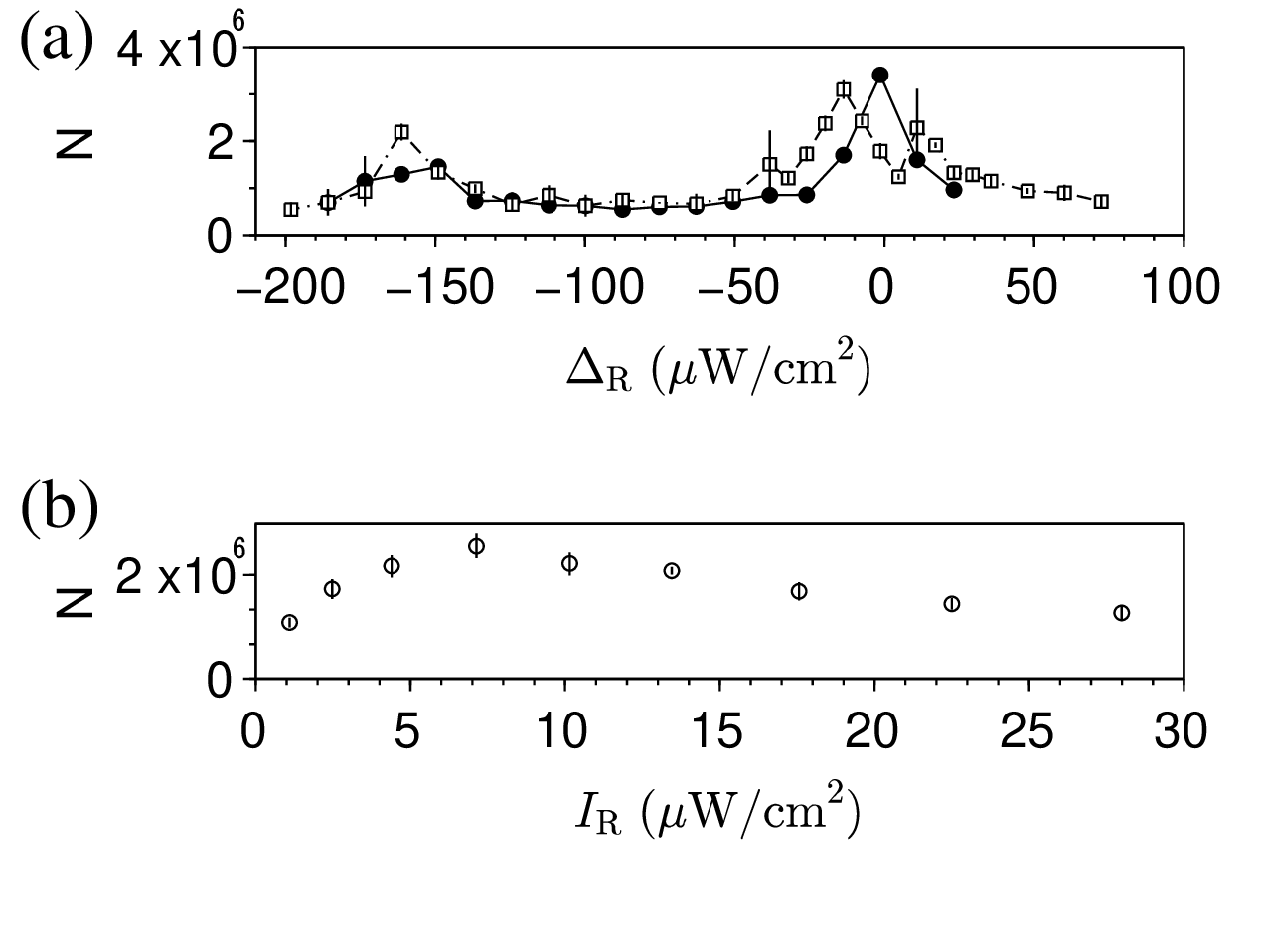}
 \caption{Dependence of atom number on the repump beam parameters.
(a) Atom number as a function of the detuning from the $\textit{F}$=1 - $\textit{F}'$ =2 resonance frequency.
The data are taken for $I_\mathrm{R}$ = 10 $\mu$W/cm$^2$ (closed circles)
and 28 $\mu$W/cm$^2$ (open squares).
The lines are guides to the eye.
(b) Dependence of atom number on the repump beam intensity 
at a fixed detuning of $\Delta_\mathrm{R}$ = $0$ MHz.
The cooling beam intensity and detuning are
$I_\mathrm{M}$ = 28 mW/cm$^2$ and $\Delta_\mathrm{M}$ = $-232$ MHz,
respectively.}
 \label{fig:repump-dependence}
\end{figure}

The atom number in the optical trap also
depends on the repump beam intensity $I_\mathrm{R}$ and detuning $\Delta_\mathrm{R}$.
Figure \ref{fig:repump-dependence} shows the atom numbers in the optical trap
for different repump parameters.
As in the case of loading from the MOT \cite{PhysRevA.62.013406},
the atom number is maximized 
when the repump beam parameters $I_\mathrm{R}$ is weak with
$\Delta_\mathrm{R}$ close to zero.
As shown in Fig. \ref{fig:repump-dependence}(a),
the profile is sharp for weak repump power of $I_\mathrm{R}$ = 10 $\mu$W/cm$^2$,
while the profile broadens and a dip around the resonance appears 
for a beam of $I_\mathrm{R}$ = 28 $\mu$W/cm$^2$
with the peak atom number slightly decreased.
This result suggests that an excessive repumping rate leads to the atom loss.
The decrease in the atom number with the increase of $I_\mathrm{R}$
is most probably related to the light-assisted collisional loss, as discussed in detail in \cite{PhysRevA.62.013406}. 
The atom number is maximum for weak and resonant
($\Delta_\mathrm{R}$ = 0 MHz) in the configuration of this experiment.

We also investigate the optimal intensity at a fixed frequency of $\Delta_\mathrm{R}$ = $0$ MHz.
The result is shown in Fig. \ref{fig:repump-dependence}(b).
The optimal beam intensity is $\approx$ 7 $\mu$W/cm$^2$. 
The atom number decreases for increasing $I_\mathrm{R}$, 
as expected from the result in Fig. \ref{fig:repump-dependence}(a).
For low intensities,
the atoms in the molasses are observed to fall because of gravity.
We suppose this sets the limit on the minimum repump intensity in our setup.

\section{Magnetic field dependence}
\begin{figure}[t]
 \centering
  \includegraphics[width=7cm]{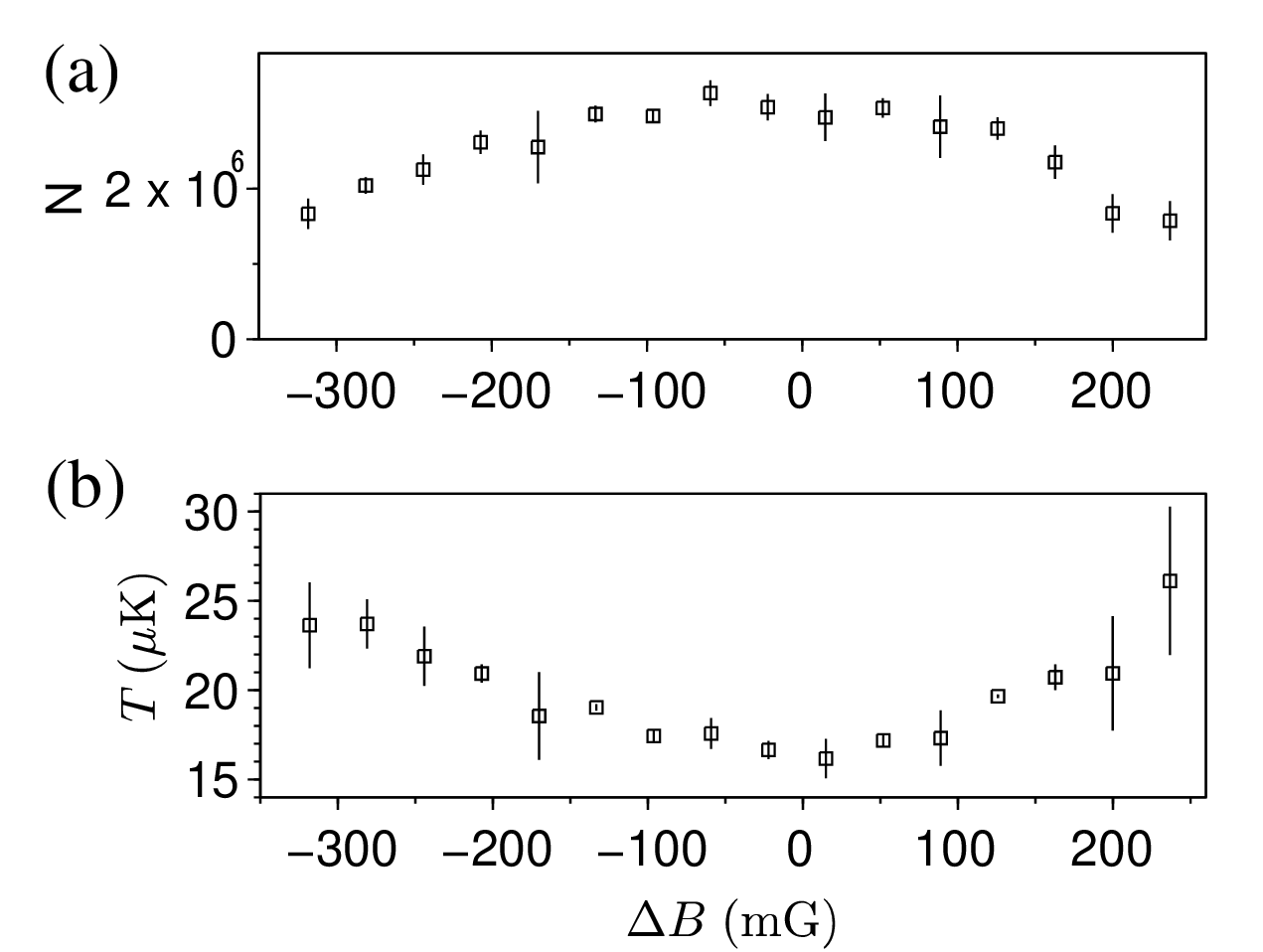}
 \caption{Influence of the magnetic field on the loading.
(a) Dependence of atom number on the added magnetic field $\Delta B$.
(b) Dependence of temperature on $\Delta B$.}
 \label{fig:B-dependence}
\end{figure}

The absence of the magnetic field is essential for the loading from the molasses.
To investigate the influence of the magnetic-field  on the loading,
we add a magnetic field $\Delta B$ during the loading
and observe the changes in the atom number and the temperature.
We regard the magnetic field giving the lowest molasses temperature
as the zero point of the magnetic field ($\Delta B$ = 0).
When adding a magnetic field $\Delta B$,
we observe the decrease in the loaded atom number 
as shown in Fig. \ref{fig:B-dependence}(a).
We ascribe this to the diffusion or heating of the molasses in the presence of the magnetic field.
In fact, the diffusion of atoms in the molasses is observed 
when the residual field is not sufficiently reduced.
The diffusion decreases the density of the molasses as a reservoir of atoms
and leads to less loaded atom number. 
The maximum atom number is obtained when $\vert \Delta B \vert < 100$ mG 
and the atom number becomes half of the maximum when 
$\vert \Delta B \vert $ is approximately 250 mG.
We also observe the temperature of the loaded atoms increases by 5 $\mu$K
when $\vert \Delta B \vert $ is 250 mG.

\begin{figure}[t]
 \centering
  \includegraphics[width=7cm]{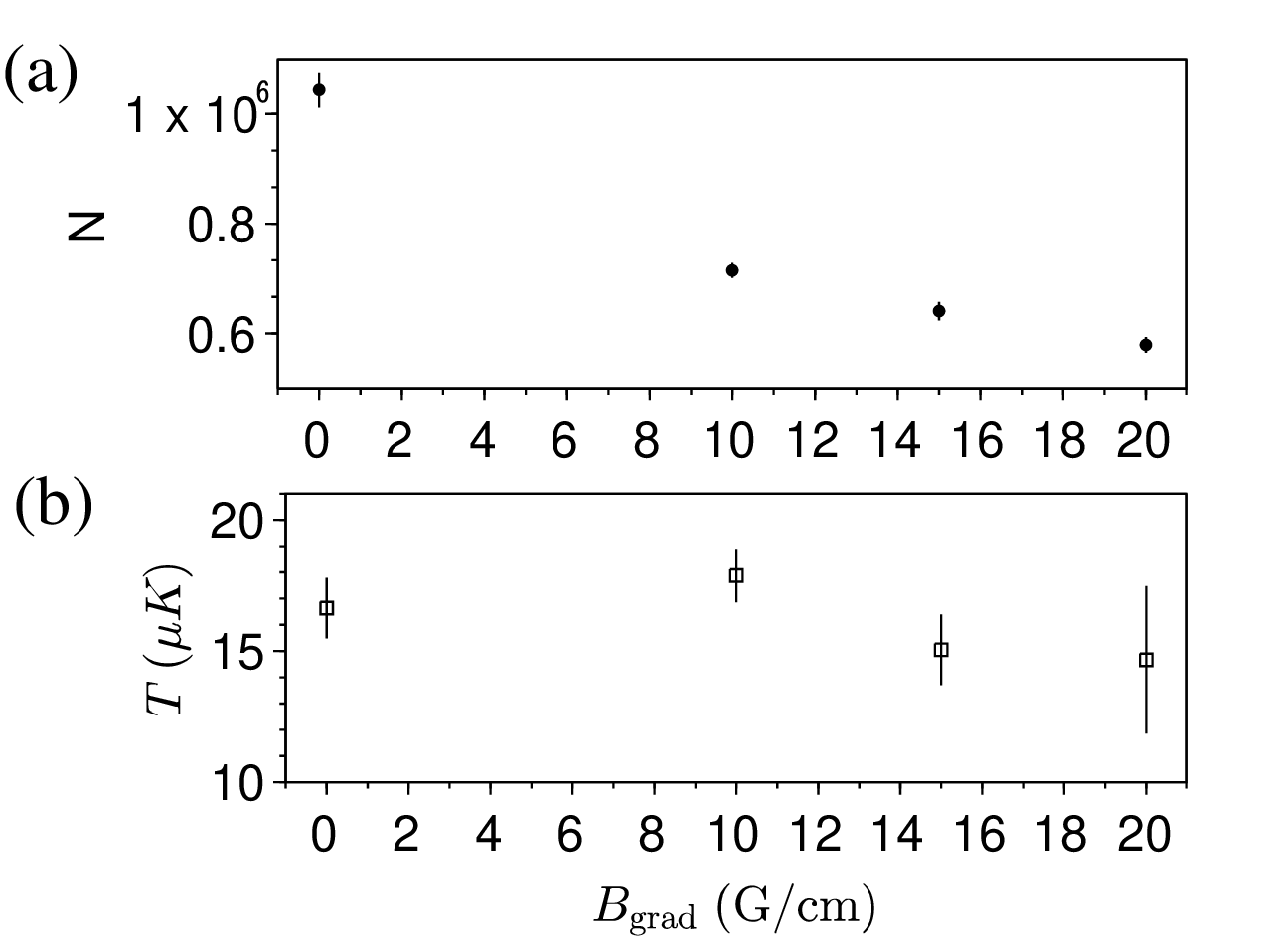}
 \caption{Influence of the magnetic field gradient on the loading.
(a) Dependence of atom number on the added magnetic field gradient $B_{\mathrm{grad}}$.
(b) Dependence of temperature on $B_{\mathrm{grad}}$.
$B_{\mathrm{grad}}$ represents the magnetic field gradient along the coil axis.}
 \label{fig:Bgrad}
\end{figure}

We also confirm the effect of magnetic field gradient on the loading efficiency.
Atom number and temperature in the optical trap are measured
when a quadrupole field generated by the coil for MOT is applied 
in the second half of the loading period.
We note that the loading parameters are different from 
that used in taking data in Fig. \ref{fig:B-dependence}.
The (total) loading period is 120 ms and the quadrupole field is applied for 60 ms.
$\Delta_\mathrm{M}$ is $-200$ MHz in this experiment.
The optical trap beam detuning from the D$_2$ line is $-$1.2 THz.
The beam power is set to 220 mW during the loading 
and decreased to 68 mW during the holding time of 25 ms after the loading
to reduce the heating by the trap beam (described later). 
Using absorption images of atoms, we confirm that the optical trap beam center is almost 
at the center of the MOT after the compression,
which is expected to coincide with the center of the quadrupole field,
to keep the magnetic field induced by the quadrupole field small.

Figure \ref{fig:Bgrad}(a) indicates that
the atom number is reduced when the magnetic field gradient is applied.
The decrease in the atom number can be explained as follows.
Because the cooling beam is very detuned from 
the cycling $\textit{F}$=2 - $\textit{F}'$ =3 transition, the MOT does not work.
The magnetic field gradient rather diffuses atoms of the reservoir,
which leads to the smaller loading rate and fewer atoms in the optical trap.
No significant change in the temperature is observed 
in this experiment as shown in Fig. \ref{fig:Bgrad}(b).

The results in Fig. \ref{fig:Bgrad} imply the efficient loading is not achieved 
with using the MOT requiring a magnetic field gradient.
The detuning of the cooling beam cannot be taken so large 
with maintaining the trapping force by MOT as discussed above.
While the loading from the molasses is optimized for the frequency close to $\textit{F}$=2 - $\textit{F}'$ =2 transition
and at the blue side of the transition, 
atoms would diffuse in the MOT configuration with such a largely detuned cooling beam.
The efficient loading in our scheme can be regarded as the result of 
the combination of large detuning and no magnetic field gradient.

\section{Loading dynamics}
\begin{figure}[t]
 \centering
 \includegraphics[width=8cm]{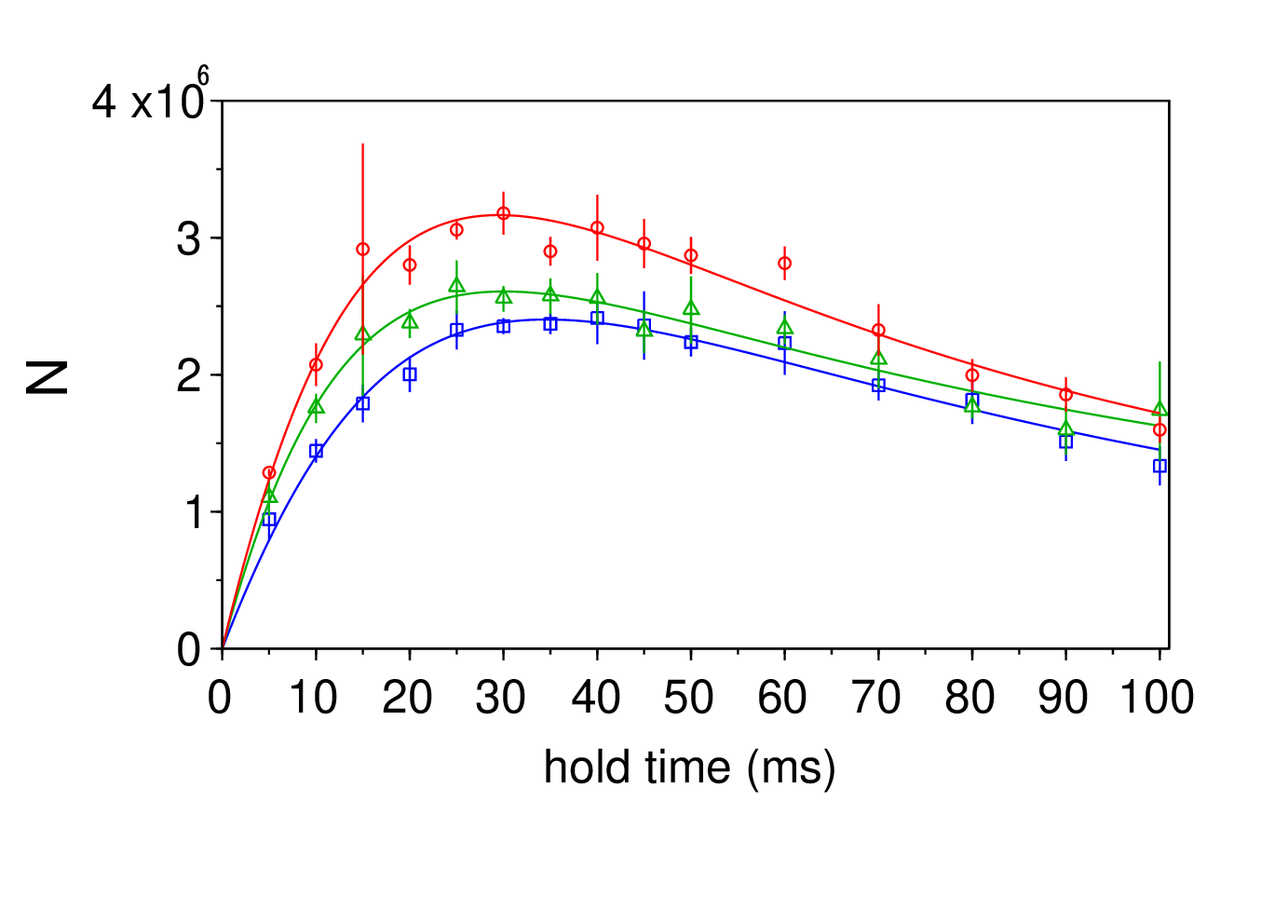}
 \caption{(color online) 
 Number of atoms in the optical trap as a function of the loading time.
 The atom number is measured for trap beam powers of
 196 mW (red circles), 151 mW (green triangles), and 105 mW (blue squares).
 The corresponding trap depths are 280 $\mu$K, 210 $\mu$K, and 150 $\mu$K, respectively.
 The solid lines are fitting curves of Eq. (\ref{rate eq}).}
 \label{fig:time-dependence}
\end{figure}

Figure \ref{fig:time-dependence} shows the loading curves for three different trap depths. 
The trap depths (and the corresponding trap beam powers) 
are 280 $\mu$K (196 mW), 210 $\mu$K (151 mW), and 150 $\mu$K (105 mW).
The cooling and repump beam parameters are
$I_\mathrm{M}$ = 28 mW/cm$^2$, $\Delta_\mathrm{M}$ = $-232$ MHz, 
$I_\mathrm{R}$ = 7.1 $\mu$W/cm$^2$, and $\Delta_\mathrm{R}$ = $0$ MHz.
As the atom number reaches a few millions within 20 ms,  
the loading rate is roughly estimated to be higher than $10^8$ atoms/s.
The solid lines are fitting curves based on the simple loading model 
described below.
The gradual decrease in the atom number is attributed to
the molasses lifetime and collisional losses.
The large maximum atom number indicates that
the loading rate exceeds the molasses loss rate and the collisional loss rate in the trap.
Although the molasses loss rate is higher than that of MOT and
one might expect the loading from the molasses is less effective,
the high loading rate enables us to load 
several million atoms from the molasses into the trap.

\begin{figure}[t]
 \centering
  \includegraphics[width=7cm]{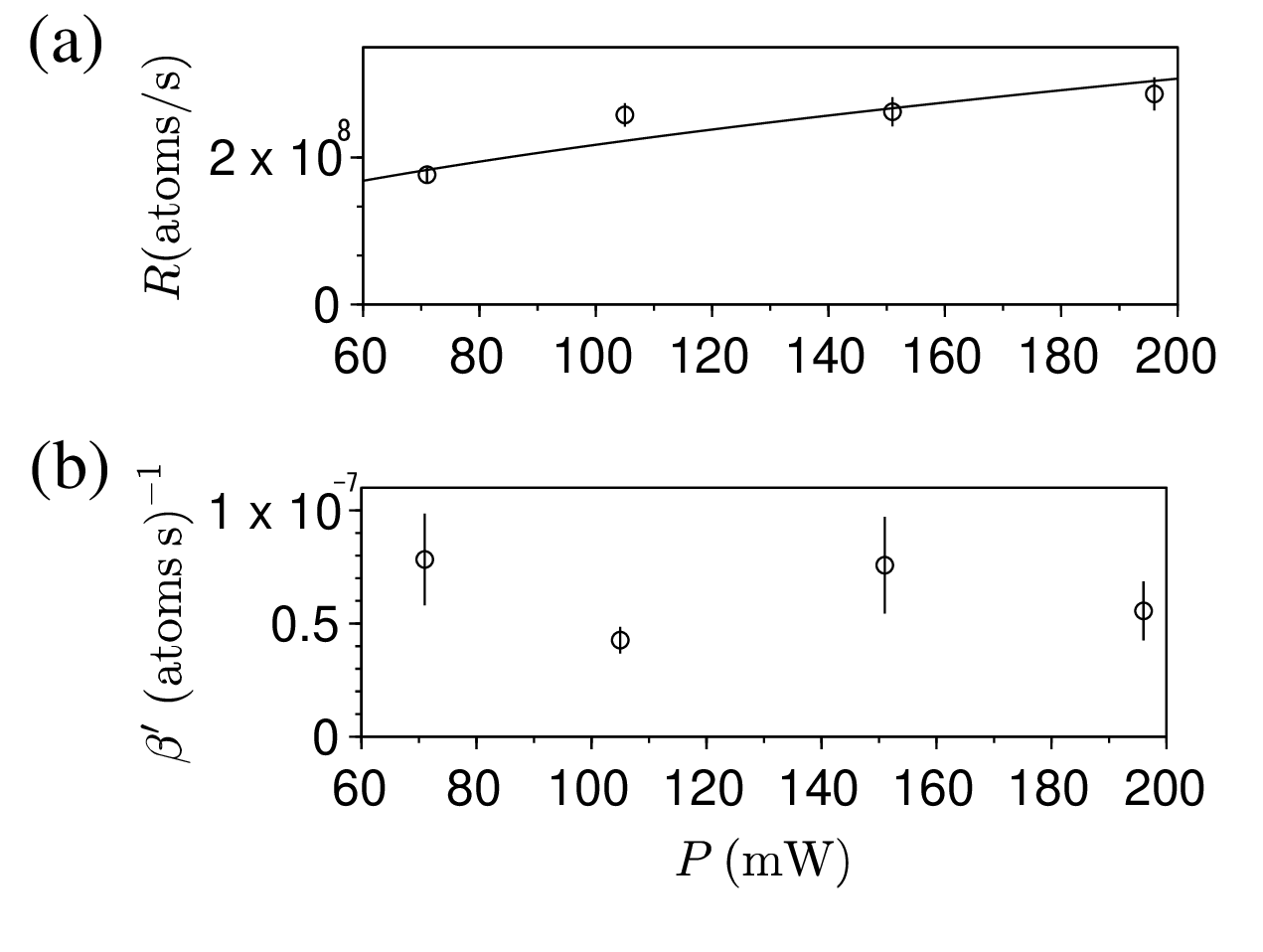}
 \caption{Loading rate (a) and two-body loss coefficient (b) 
plotted against the trapping beam power $P$.
The solid line in (a) is the fit by $R=A\sqrt{P}$, 
where $A = 2.2 (1) \times 10^6$ atoms/(s $\sqrt{\mathrm{mW} }$ ).
}
 \label{fig:Randb}
\end{figure}

The time dependence of the atom number in the optical trap $N$ can be described by
\begin{equation}\label{rate eq}
\frac{dN}{dt} = R_0 \exp ( -\gamma_\mathrm{M}t) - \beta' N^2,
\end{equation}
where
$R_0$ is the initial loading rate and
$\gamma_\mathrm{M}$ and $\beta'$ are the loss rate of the molasses 
and the two-body loss coefficient, respectively \cite{PhysRevA.62.013406}.
We omit the one-body loss term because it is negligible in the concerned time scale. 
We also neglect the three-body collisional loss term for the simplicity of the discussion.
This neglect is safe because 
the two-body loss (which is due to light-assisted collision by the cooling beam
as discussed below) is dominant over the three-body collisional loss
in the loading process considered here.
The model in Eq.(\ref{rate eq}) fits the experimental data.
The fitted parameters are plotted as a function of the trap beam power $P$
in Fig. \ref{fig:Randb}.
As shown in Fig. \ref{fig:Randb}(a),
the loading rate $R_0$ ranges from $1.8 (1) \times 10^8$ atoms/s to $2.9 (2) \times 10^8$ atoms/s.
The rates are almost 10 times higher
than that achieved by loading directly from the MOT, 
in which case $R_0$ is approximately $3 \times 10^7$ atoms/s when the trap depth is $\sim$ 1 mK 
\cite{PhysRevA.62.013406}.

The high loading rate is considered to be resulted from the high density of molasses.
Since $nv$ ($n:$ atom density, $v:$ the atom velocity) is almost constant in the sub-Doppler regime \cite{PhysRevA.52.1423},
low temperature of the molasses leads to high density.
Although the loading rate depends on atom flux into the optical trap
and the flux is proportional to $nv$,
we suspect that the $v$ of an atom entering the trap 
is determined by the trap potential depth $U_0$, independent of $n$.
Because the molasses temperature is much less than $U_0/k_\mathrm{B}$,
the conversion of a fraction of $U_0$ to kinetic energy
determines the $v$ of an atom entering the trap.
The fact that the observed loading rate is proportional to $\sqrt{P}$, 
as shown in Fig. \ref{fig:Randb}(a), supports 
the inference that a fraction of $U_0$ is converted to $v$. 
The high damping coefficient in PGC 
is another possible reason for the increase of the loading rate.
As a large damping coefficient leads to the overdamping of atom motion in the trap,
an atom entering the trap is expected to be trapped with a high trapping probability.

The values of the fitted two-body loss coefficients shown in Fig. \ref{fig:Randb}(b)
are comparable to the results in \cite{PhysRevA.62.013406},
in which $\beta'$ of several $10^{-6}$ (atoms s) $^{-1}$ is reported 
for a trap beam of $w_0$ = 26 $\mu$m.
As already discussed in \cite{PhysRevA.62.013406},
the two-body loss is mainly caused by light-assisted collisions.
The molasses loss rate $\gamma_\mathrm{M}$ is 
around 50 s$^{-1}$ for any $P$,
which is consistent with the decrease of the number of atoms in the molasses.

\section{Initial temperature and phase-space density}


In order to measure the initial temperature $T_0$ immediately after the completion of loading,
the heating due to the photon scattering of the trap beam during the holding time
should be taken into account
because the small detuning of $-2.3$ THz from the D$_2$ line
leads to a high photon scattering rate
in comparison with large detuning in far-off resonant trap.
We measure the heating rate in the optical trap  
as 0.30 (5) $\mu$K/(s mW) from the temperatures measured after several holding durations from 25 ms to 95 ms assuming a constant heating rate.

For the optimal parameters
($I_\mathrm{M}$ = 28 mW/cm$^2$, $\Delta_\mathrm{M}$ = $-232$ MHz, 
$I_\mathrm{R}$ = 7.1 $\mu$W/cm$^2$, $\Delta_\mathrm{R}$ = $0$ MHz
and 40 ms loading),
we produce atom clouds in the optical trap  
with the atom number $N$ and temperature $T$ after the holding of 25 ms
being $3.1 (2) \times 10^6$ and $16.2 (5)$ $\mu$K, respectively.
Using the measured heating rate, 
we estimate the initial temperature $T_0$ to be 14.7 (7) $\mu$K
corresponding to $\sim U/20$,
although $T_0$ is higher than the temperature of atoms in the molasses.

The initial low temperature in the optical trap of tight confinement 
naturally leads to a high PSD.
For the above case,
the initial PSD estimated from $N$ and $T_0$ is $1.1 \times 10^{-3}$.
This is one or two orders of magnitude greater than that in the loading from the MOT, which typically gives $\rho \sim 10^{-4}$ or $10^{-5}$. 
It is noted that the true initial PSD is greater than this estimation
because the initial atom number $N_0$ is reduced to $N$ during the holding time.

In order to take advantage of this high initial PSD, 
it is desirable to suppress heating and atom loss after loading.
While light-assisted two-body loss diminishes after loading with the optical molasses beams finishes, we need to consider the three-body collisional loss. 
In particular, the cold atoms in the trap of tight confinement
with a high peak density ($\propto \bar{\nu}^3 T^{-3/2}$, $\bar{\nu}$ : mean trap frequency) are susceptible to the three-body collisional loss.
While the peak density in the deepest trap in this work
is estimated to be $9 \times 10^{12}$ cm$^{-3}$,
higher density near $10^{14}$ cm$^{-3}$ leads to a non-negligible three-body collisional loss rate \cite{PhysRevLett.79.337}. 
We find that the sudden decrease in the trap beam power after the loading 
can be a countermeasure against the high density.
By reducing the trap beam power to about half of the initial value immediately after the loading, the temperature decreases with the atom number 
almost same to that without the trapping beam power reduction.
Although the PSD decreases slightly,
this is a good start point for the evaporation
because the loss and heating of atoms are suppressed in the trap of reduced power.

\section{Discussion}\label{sec: discussion}
We discuss the characteristic features of the loading scheme in this work compared with other loading schemes.
First, as commented already, the loading from the molasses 
can reach to lower temperature than the loading from the MOT.
The loading is sensitive to the magnetic field as shown in Fig. \ref{fig:B-dependence}.
Next, we compare our scheme with other works using the optical molasses
for the optical trap loading.
There have been few reports on the loading with the optical molasses.
Dumke \textit{et al.} \cite{Dumke2006} used transitional molasses phase where 
the magnetic-field gradient for the MOT is gradually decreased in 60 ms
for the loading of $^{23} \mathrm{Na}$ atoms into the crossed beam trap.
The reported initial temperature is approximately 1/10 of the potential depths
and the effect of PGC is unclear.
Recently, the efficient loading using $D_1$ gray-molasses is realized
for $^{39}\mathrm{K}$ \cite{PhysRevA.90.033405} and $^{6}\mathrm{Li}$ \cite{Burchianti2014}. 
For these atoms, the small hyperfine splitting prevents effective PGC in the MOT and the use of $D_1$ gray-molasses is a crude choice.
Nevertheless, the loading using normal (``bright'') molasses on $D_2$ line for the MOT
has not been reported as far as we know, 
although this loading method seem to be simple (for atoms in which bright molasses effectively works) to be implemented.
The required large detuning of the cooling beam ($\Delta_{\mathrm{M}}$) for optimizing the loading
might have been making the pursuit of the loading from the molasses difficult.
The result in Fig. \ref{fig:cooling-dependence} indicates that the increase in the atom number is not due to the cooling by the large detuning.
We ascribe the increase of the atom number to
small light-assisted collisional when the detuning is large.
The analysis of the loading dynamics for different $\Delta_{\mathrm{M}}$ shows that
$\beta'$ decreases when the $\Delta_{\mathrm{M}}$ is changed from $-200$ MHz to $-248$ MHz,
while $R_0$ is almost constant.

\section{Summary and Outlook}\label{sec: summary}
We successfully loaded a large number of atoms with low temperature
in an optical trap by simply overlapping molasses and the trap beam.
The loading is optimized when the beam detuning due to molasses is approximately $-230$ MHz,
and the loaded atoms are cooled to a temperature less than 1/10 of the trap depth. 
Despite the short molasses lifetime, 
the optimized loading enables the molasses atoms
to be loaded into the optical trap efficiently within the lifetime of the molasses.

The high initial PSD is advantageous for reducing the evaporation time.
A short evaporation time is helpful for experiments requiring a high repetition rate. 
The reduction in the evaporation time also decreases the loss during the evaporation,
including the loss due to collision with the hot background atoms.
Thus, the high initial PSD will enable one to create a quantum gas 
in a rough vacuum environment.
This is particularly suitable for hybrid experiments combining cold atoms with other systems.
The fast evaporation might also open the possibility
of evaporative cooling in an optical trap of a moderate detuning, as was used in this work.

A larger atom number is also expected when using a trap having a larger volume
combined with the loading from the molasses.
Loading of more than $1 \times 10^7$ atoms is expected 
with the use of a high power and wide trap beam,
as demonstrated with the gray-molasses \cite{PhysRevA.90.033405, Burchianti2014}. 
The expected stagnation after the evaporation is proceeded
due to the low density and collisional rate between atoms in a large trap
can be compensated by an elaborate dynamical beam-size change 
\cite{PhysRevA.93.043403,PhysRevA.95.013609}.
Combining these techniques, the all-optical production of a BEC with $1 \times 10^6$ or even more atoms is expected.
Such a large BEC in an optical trap is suitable for many experiments
including multi-component and spinor BEC experiments \cite{RevModPhys.85.1191, PhysRevLett.81.1539, Stenger1998, PhysRevLett.92.140403}.

The loading method demonstrated here should be applicable to other 
experiments of cold atoms, and atoms with large hyperfine splitting in particular,
because no additional lasers nor changes in the MOT paths are required.
The method will broaden the applications of cold atom gases
owing to its capability of producing ultracold or quantum degenerate gases
of large number of atoms in a short time. 

\begin{acknowledgements}
We would like to thank T. Hibino, N. Ichinoseki, and S. Fujita 
for their assistance in constructing the earliest experimental setup. 
This work was supported by the Matsuo Foundation; the Research Foundation for Opto-Science and Technology; Grant-in-Aid for Scientific Research (Nos. 26887033, 15K05234, 23740308) from the Ministry of Education, Culture, Sports, Science, and Technology of Japan; and Chuo University Joint Research Grant.
\end{acknowledgements}


\begin{thebibliography}{26}
\expandafter\ifx\csname natexlab\endcsname\relax\def\natexlab#1{#1}\fi
\expandafter\ifx\csname bibnamefont\endcsname\relax
  \def\bibnamefont#1{#1}\fi
\expandafter\ifx\csname bibfnamefont\endcsname\relax
  \def\bibfnamefont#1{#1}\fi
\expandafter\ifx\csname citenamefont\endcsname\relax
  \def\citenamefont#1{#1}\fi
\expandafter\ifx\csname url\endcsname\relax
  \def\url#1{\texttt{#1}}\fi
\expandafter\ifx\csname urlprefix\endcsname\relax\def\urlprefix{URL }\fi
\providecommand{\bibinfo}[2]{#2}
\providecommand{\eprint}[2][]{\url{#2}}

\bibitem[{\citenamefont{Grimm et~al.}(2000)\citenamefont{Grimm, Weidemuller,
  and Ovchinnikov}}]{Grimm200095}
\bibinfo{author}{\bibfnamefont{R.}~\bibnamefont{Grimm}},
  \bibinfo{author}{\bibfnamefont{M.}~\bibnamefont{Weidemuller}},
  \bibnamefont{and} \bibinfo{author}{\bibfnamefont{Y.~B.}
  \bibnamefont{Ovchinnikov}}, \bibinfo{journal}{Advances In Atomic, Molecular,
  and Optical Physics} \textbf{\bibinfo{volume}{42}}, \bibinfo{pages}{95 }
  (\bibinfo{year}{2000}).

\bibitem[{\citenamefont{Tojo et~al.}(2009)\citenamefont{Tojo, Hayashi, Tanabe,
  Hirano, Kawaguchi, Saito, and Ueda}}]{PhysRevA.80.042704}
\bibinfo{author}{\bibfnamefont{S.}~\bibnamefont{Tojo}},
  \bibinfo{author}{\bibfnamefont{T.}~\bibnamefont{Hayashi}},
  \bibinfo{author}{\bibfnamefont{T.}~\bibnamefont{Tanabe}},
  \bibinfo{author}{\bibfnamefont{T.}~\bibnamefont{Hirano}},
  \bibinfo{author}{\bibfnamefont{Y.}~\bibnamefont{Kawaguchi}},
  \bibinfo{author}{\bibfnamefont{H.}~\bibnamefont{Saito}}, \bibnamefont{and}
  \bibinfo{author}{\bibfnamefont{M.}~\bibnamefont{Ueda}},
  \bibinfo{journal}{Phys. Rev. A} \textbf{\bibinfo{volume}{80}},
  \bibinfo{pages}{042704} (\bibinfo{year}{2009}).

\bibitem[{\citenamefont{Stamper-Kurn and Ueda}(2013)}]{RevModPhys.85.1191}
\bibinfo{author}{\bibfnamefont{D.~M.} \bibnamefont{Stamper-Kurn}}
  \bibnamefont{and} \bibinfo{author}{\bibfnamefont{M.}~\bibnamefont{Ueda}},
  \bibinfo{journal}{Rev. Mod. Phys.} \textbf{\bibinfo{volume}{85}},
  \bibinfo{pages}{1191} (\bibinfo{year}{2013}).

\bibitem[{\citenamefont{Hall et~al.}(1998)\citenamefont{Hall, Matthews, Ensher,
  Wieman, and Cornell}}]{PhysRevLett.81.1539}
\bibinfo{author}{\bibfnamefont{D.~S.} \bibnamefont{Hall}},
  \bibinfo{author}{\bibfnamefont{M.~R.} \bibnamefont{Matthews}},
  \bibinfo{author}{\bibfnamefont{J.~R.} \bibnamefont{Ensher}},
  \bibinfo{author}{\bibfnamefont{C.~E.} \bibnamefont{Wieman}},
  \bibnamefont{and} \bibinfo{author}{\bibfnamefont{E.~A.}
  \bibnamefont{Cornell}}, \bibinfo{journal}{Phys. Rev. Lett.}
  \textbf{\bibinfo{volume}{81}}, \bibinfo{pages}{1539} (\bibinfo{year}{1998}).

\bibitem[{\citenamefont{Stenger et~al.}(1998)\citenamefont{Stenger, Inouye,
  Stamper-Kurn, Miesner, Chikkatur, and Ketterle}}]{Stenger1998}
\bibinfo{author}{\bibfnamefont{J.}~\bibnamefont{Stenger}},
  \bibinfo{author}{\bibfnamefont{S.}~\bibnamefont{Inouye}},
  \bibinfo{author}{\bibfnamefont{D.~M.} \bibnamefont{Stamper-Kurn}},
  \bibinfo{author}{\bibfnamefont{H.-J.} \bibnamefont{Miesner}},
  \bibinfo{author}{\bibfnamefont{A.~P.} \bibnamefont{Chikkatur}},
  \bibnamefont{and} \bibinfo{author}{\bibfnamefont{W.}~\bibnamefont{Ketterle}},
  \bibinfo{journal}{Nature} \textbf{\bibinfo{volume}{396}},
  \bibinfo{pages}{345} (\bibinfo{year}{1998}).

\bibitem[{\citenamefont{Chang et~al.}(2004)\citenamefont{Chang, Hamley,
  Barrett, Sauer, Fortier, Zhang, You, and Chapman}}]{PhysRevLett.92.140403}
\bibinfo{author}{\bibfnamefont{M.-S.} \bibnamefont{Chang}},
  \bibinfo{author}{\bibfnamefont{C.~D.} \bibnamefont{Hamley}},
  \bibinfo{author}{\bibfnamefont{M.~D.} \bibnamefont{Barrett}},
  \bibinfo{author}{\bibfnamefont{J.~A.} \bibnamefont{Sauer}},
  \bibinfo{author}{\bibfnamefont{K.~M.} \bibnamefont{Fortier}},
  \bibinfo{author}{\bibfnamefont{W.}~\bibnamefont{Zhang}},
  \bibinfo{author}{\bibfnamefont{L.}~\bibnamefont{You}}, \bibnamefont{and}
  \bibinfo{author}{\bibfnamefont{M.~S.} \bibnamefont{Chapman}},
  \bibinfo{journal}{Phys. Rev. Lett.} \textbf{\bibinfo{volume}{92}},
  \bibinfo{pages}{140403} (\bibinfo{year}{2004}).

\bibitem[{\citenamefont{Gaunt et~al.}(2013)\citenamefont{Gaunt, Schmidutz,
  Gotlibovych, Smith, and Hadzibabic}}]{PhysRevLett.110.200406}
\bibinfo{author}{\bibfnamefont{A.~L.} \bibnamefont{Gaunt}},
  \bibinfo{author}{\bibfnamefont{T.~F.} \bibnamefont{Schmidutz}},
  \bibinfo{author}{\bibfnamefont{I.}~\bibnamefont{Gotlibovych}},
  \bibinfo{author}{\bibfnamefont{R.~P.} \bibnamefont{Smith}}, \bibnamefont{and}
  \bibinfo{author}{\bibfnamefont{Z.}~\bibnamefont{Hadzibabic}},
  \bibinfo{journal}{Phys. Rev. Lett.} \textbf{\bibinfo{volume}{110}},
  \bibinfo{pages}{200406} (\bibinfo{year}{2013}).

\bibitem[{\citenamefont{Gustavson et~al.}(2001)\citenamefont{Gustavson,
  Chikkatur, Leanhardt, G\"orlitz, Gupta, Pritchard, and
  Ketterle}}]{PhysRevLett.88.020401}
\bibinfo{author}{\bibfnamefont{T.~L.} \bibnamefont{Gustavson}},
  \bibinfo{author}{\bibfnamefont{A.~P.} \bibnamefont{Chikkatur}},
  \bibinfo{author}{\bibfnamefont{A.~E.} \bibnamefont{Leanhardt}},
  \bibinfo{author}{\bibfnamefont{A.}~\bibnamefont{G\"orlitz}},
  \bibinfo{author}{\bibfnamefont{S.}~\bibnamefont{Gupta}},
  \bibinfo{author}{\bibfnamefont{D.~E.} \bibnamefont{Pritchard}},
  \bibnamefont{and} \bibinfo{author}{\bibfnamefont{W.}~\bibnamefont{Ketterle}},
  \bibinfo{journal}{Phys. Rev. Lett.} \textbf{\bibinfo{volume}{88}},
  \bibinfo{pages}{020401} (\bibinfo{year}{2001}).

\bibitem[{\citenamefont{Roy et~al.}(2016)\citenamefont{Roy, Green, Bowler, and
  Gupta}}]{PhysRevA.93.043403}
\bibinfo{author}{\bibfnamefont{R.}~\bibnamefont{Roy}},
  \bibinfo{author}{\bibfnamefont{A.}~\bibnamefont{Green}},
  \bibinfo{author}{\bibfnamefont{R.}~\bibnamefont{Bowler}}, \bibnamefont{and}
  \bibinfo{author}{\bibfnamefont{S.}~\bibnamefont{Gupta}},
  \bibinfo{journal}{Phys. Rev. A} \textbf{\bibinfo{volume}{93}},
  \bibinfo{pages}{043403} (\bibinfo{year}{2016}).

\bibitem[{\citenamefont{L\'eonard et~al.}(2014)\citenamefont{L\'eonard, Lee,
  Morales, Karg, Esslinger, and Donner}}]{1367-2630-16-9-093028}
\bibinfo{author}{\bibfnamefont{J.}~\bibnamefont{L\'eonard}},
  \bibinfo{author}{\bibfnamefont{M.}~\bibnamefont{Lee}},
  \bibinfo{author}{\bibfnamefont{A.}~\bibnamefont{Morales}},
  \bibinfo{author}{\bibfnamefont{T.~M.} \bibnamefont{Karg}},
  \bibinfo{author}{\bibfnamefont{T.}~\bibnamefont{Esslinger}},
  \bibnamefont{and} \bibinfo{author}{\bibfnamefont{T.}~\bibnamefont{Donner}},
  \bibinfo{journal}{New Journal of Physics} \textbf{\bibinfo{volume}{16}},
  \bibinfo{pages}{093028} (\bibinfo{year}{2014}).

\bibitem[{\citenamefont{Takasu et~al.}(2003)\citenamefont{Takasu, Maki, Komori,
  Takano, Honda, Kumakura, Yabuzaki, and Takahashi}}]{PhysRevLett.91.040404}
\bibinfo{author}{\bibfnamefont{Y.}~\bibnamefont{Takasu}},
  \bibinfo{author}{\bibfnamefont{K.}~\bibnamefont{Maki}},
  \bibinfo{author}{\bibfnamefont{K.}~\bibnamefont{Komori}},
  \bibinfo{author}{\bibfnamefont{T.}~\bibnamefont{Takano}},
  \bibinfo{author}{\bibfnamefont{K.}~\bibnamefont{Honda}},
  \bibinfo{author}{\bibfnamefont{M.}~\bibnamefont{Kumakura}},
  \bibinfo{author}{\bibfnamefont{T.}~\bibnamefont{Yabuzaki}}, \bibnamefont{and}
  \bibinfo{author}{\bibfnamefont{Y.}~\bibnamefont{Takahashi}},
  \bibinfo{journal}{Phys. Rev. Lett.} \textbf{\bibinfo{volume}{91}},
  \bibinfo{pages}{040404} (\bibinfo{year}{2003}).

\bibitem[{\citenamefont{Griesmaier et~al.}(2005)\citenamefont{Griesmaier,
  Werner, Hensler, Stuhler, and Pfau}}]{PhysRevLett.94.160401}
\bibinfo{author}{\bibfnamefont{A.}~\bibnamefont{Griesmaier}},
  \bibinfo{author}{\bibfnamefont{J.}~\bibnamefont{Werner}},
  \bibinfo{author}{\bibfnamefont{S.}~\bibnamefont{Hensler}},
  \bibinfo{author}{\bibfnamefont{J.}~\bibnamefont{Stuhler}}, \bibnamefont{and}
  \bibinfo{author}{\bibfnamefont{T.}~\bibnamefont{Pfau}},
  \bibinfo{journal}{Phys. Rev. Lett.} \textbf{\bibinfo{volume}{94}},
  \bibinfo{pages}{160401} (\bibinfo{year}{2005}).

\bibitem[{\citenamefont{Barrett et~al.}(2001)\citenamefont{Barrett, Sauer, and
  Chapman}}]{PhysRevLett.87.010404}
\bibinfo{author}{\bibfnamefont{M.~D.} \bibnamefont{Barrett}},
  \bibinfo{author}{\bibfnamefont{J.~A.} \bibnamefont{Sauer}}, \bibnamefont{and}
  \bibinfo{author}{\bibfnamefont{M.~S.} \bibnamefont{Chapman}},
  \bibinfo{journal}{Phys. Rev. Lett.} \textbf{\bibinfo{volume}{87}},
  \bibinfo{pages}{010404} (\bibinfo{year}{2001}).

\bibitem[{\citenamefont{Arnold and Barrett}(2011)}]{Arnold20113288}
\bibinfo{author}{\bibfnamefont{K.}~\bibnamefont{Arnold}} \bibnamefont{and}
  \bibinfo{author}{\bibfnamefont{M.}~\bibnamefont{Barrett}},
  \bibinfo{journal}{Optics Communications} \textbf{\bibinfo{volume}{284}},
  \bibinfo{pages}{3288 } (\bibinfo{year}{2011}).

\bibitem[{\citenamefont{Kumar et~al.}(2012)\citenamefont{Kumar, Hirai, Suzuki,
  Kachi, Sadgrove, and Nakagawa}}]{JPSJ.81.084004}
\bibinfo{author}{\bibfnamefont{S.}~\bibnamefont{Kumar}},
  \bibinfo{author}{\bibfnamefont{S.}~\bibnamefont{Hirai}},
  \bibinfo{author}{\bibfnamefont{Y.}~\bibnamefont{Suzuki}},
  \bibinfo{author}{\bibfnamefont{M.}~\bibnamefont{Kachi}},
  \bibinfo{author}{\bibfnamefont{M.}~\bibnamefont{Sadgrove}}, \bibnamefont{and}
  \bibinfo{author}{\bibfnamefont{K.}~\bibnamefont{Nakagawa}},
  \bibinfo{journal}{Journal of the Physical Society of Japan}
  \textbf{\bibinfo{volume}{81}}, \bibinfo{pages}{084004}
  (\bibinfo{year}{2012}).

\bibitem[{\citenamefont{Kinoshita et~al.}(2005)\citenamefont{Kinoshita, Wenger,
  and Weiss}}]{PhysRevA.71.011602}
\bibinfo{author}{\bibfnamefont{T.}~\bibnamefont{Kinoshita}},
  \bibinfo{author}{\bibfnamefont{T.}~\bibnamefont{Wenger}}, \bibnamefont{and}
  \bibinfo{author}{\bibfnamefont{D.~S.} \bibnamefont{Weiss}},
  \bibinfo{journal}{Phys. Rev. A} \textbf{\bibinfo{volume}{71}},
  \bibinfo{pages}{011602} (\bibinfo{year}{2005}).

\bibitem[{\citenamefont{Hemmerich et~al.}(1995)\citenamefont{Hemmerich,
  Weidem\"uller, Esslinger, Zimmermann, and H\"ansch}}]{PhysRevLett.75.37}
\bibinfo{author}{\bibfnamefont{A.}~\bibnamefont{Hemmerich}},
  \bibinfo{author}{\bibfnamefont{M.}~\bibnamefont{Weidem\"uller}},
  \bibinfo{author}{\bibfnamefont{T.}~\bibnamefont{Esslinger}},
  \bibinfo{author}{\bibfnamefont{C.}~\bibnamefont{Zimmermann}},
  \bibnamefont{and} \bibinfo{author}{\bibfnamefont{T.}~\bibnamefont{H\"ansch}},
  \bibinfo{journal}{Phys. Rev. Lett.} \textbf{\bibinfo{volume}{75}},
  \bibinfo{pages}{37} (\bibinfo{year}{1995}).

\bibitem[{\citenamefont{Boiron et~al.}(1996)\citenamefont{Boiron, Michaud,
  Lemonde, Castin, Salomon, Weyers, Szymaniec, Cognet, and
  Clairon}}]{PhysRevA.53.R3734}
\bibinfo{author}{\bibfnamefont{D.}~\bibnamefont{Boiron}},
  \bibinfo{author}{\bibfnamefont{A.}~\bibnamefont{Michaud}},
  \bibinfo{author}{\bibfnamefont{P.}~\bibnamefont{Lemonde}},
  \bibinfo{author}{\bibfnamefont{Y.}~\bibnamefont{Castin}},
  \bibinfo{author}{\bibfnamefont{C.}~\bibnamefont{Salomon}},
  \bibinfo{author}{\bibfnamefont{S.}~\bibnamefont{Weyers}},
  \bibinfo{author}{\bibfnamefont{K.}~\bibnamefont{Szymaniec}},
  \bibinfo{author}{\bibfnamefont{L.}~\bibnamefont{Cognet}}, \bibnamefont{and}
  \bibinfo{author}{\bibfnamefont{A.}~\bibnamefont{Clairon}},
  \bibinfo{journal}{Phys. Rev. A} \textbf{\bibinfo{volume}{53}},
  \bibinfo{pages}{R3734} (\bibinfo{year}{1996}).

\bibitem[{\citenamefont{Kuppens et~al.}(2000)\citenamefont{Kuppens, Corwin,
  Miller, Chupp, and Wieman}}]{PhysRevA.62.013406}
\bibinfo{author}{\bibfnamefont{S.~J.~M.} \bibnamefont{Kuppens}},
  \bibinfo{author}{\bibfnamefont{K.~L.} \bibnamefont{Corwin}},
  \bibinfo{author}{\bibfnamefont{K.~W.} \bibnamefont{Miller}},
  \bibinfo{author}{\bibfnamefont{T.~E.} \bibnamefont{Chupp}}, \bibnamefont{and}
  \bibinfo{author}{\bibfnamefont{C.~E.} \bibnamefont{Wieman}},
  \bibinfo{journal}{Phys. Rev. A} \textbf{\bibinfo{volume}{62}},
  \bibinfo{pages}{013406} (\bibinfo{year}{2000}).

\bibitem[{\citenamefont{Chu et~al.}(1986)\citenamefont{Chu, Bjorkholm, Ashkin,
  and Cable}}]{PhysRevLett.57.314}
\bibinfo{author}{\bibfnamefont{S.}~\bibnamefont{Chu}},
  \bibinfo{author}{\bibfnamefont{J.~E.} \bibnamefont{Bjorkholm}},
  \bibinfo{author}{\bibfnamefont{A.}~\bibnamefont{Ashkin}}, \bibnamefont{and}
  \bibinfo{author}{\bibfnamefont{A.}~\bibnamefont{Cable}},
  \bibinfo{journal}{Phys. Rev. Lett.} \textbf{\bibinfo{volume}{57}},
  \bibinfo{pages}{314} (\bibinfo{year}{1986}).

\bibitem[{\citenamefont{Townsend et~al.}(1995)\citenamefont{Townsend, Edwards,
  Cooper, Zetie, Foot, Steane, Szriftgiser, Perrin, and
  Dalibard}}]{PhysRevA.52.1423}
\bibinfo{author}{\bibfnamefont{C.~G.} \bibnamefont{Townsend}},
  \bibinfo{author}{\bibfnamefont{N.~H.} \bibnamefont{Edwards}},
  \bibinfo{author}{\bibfnamefont{C.~J.} \bibnamefont{Cooper}},
  \bibinfo{author}{\bibfnamefont{K.~P.} \bibnamefont{Zetie}},
  \bibinfo{author}{\bibfnamefont{C.~J.} \bibnamefont{Foot}},
  \bibinfo{author}{\bibfnamefont{A.~M.} \bibnamefont{Steane}},
  \bibinfo{author}{\bibfnamefont{P.}~\bibnamefont{Szriftgiser}},
  \bibinfo{author}{\bibfnamefont{H.}~\bibnamefont{Perrin}}, \bibnamefont{and}
  \bibinfo{author}{\bibfnamefont{J.}~\bibnamefont{Dalibard}},
  \bibinfo{journal}{Phys. Rev. A} \textbf{\bibinfo{volume}{52}},
  \bibinfo{pages}{1423} (\bibinfo{year}{1995}).

\bibitem[{\citenamefont{Burt et~al.}(1997)\citenamefont{Burt, Ghrist, Myatt,
  Holland, Cornell, and Wieman}}]{PhysRevLett.79.337}
\bibinfo{author}{\bibfnamefont{E.~A.} \bibnamefont{Burt}},
  \bibinfo{author}{\bibfnamefont{R.~W.} \bibnamefont{Ghrist}},
  \bibinfo{author}{\bibfnamefont{C.~J.} \bibnamefont{Myatt}},
  \bibinfo{author}{\bibfnamefont{M.~J.} \bibnamefont{Holland}},
  \bibinfo{author}{\bibfnamefont{E.~A.} \bibnamefont{Cornell}},
  \bibnamefont{and} \bibinfo{author}{\bibfnamefont{C.~E.}
  \bibnamefont{Wieman}}, \bibinfo{journal}{Phys. Rev. Lett.}
  \textbf{\bibinfo{volume}{79}}, \bibinfo{pages}{337} (\bibinfo{year}{1997}).

\bibitem[{\citenamefont{Dumke et~al.}(2006)\citenamefont{Dumke, Johanning,
  Gomez, Weinstein, Jones, and Lett}}]{Dumke2006}
\bibinfo{author}{\bibfnamefont{R.}~\bibnamefont{Dumke}},
  \bibinfo{author}{\bibfnamefont{M.}~\bibnamefont{Johanning}},
  \bibinfo{author}{\bibfnamefont{E.}~\bibnamefont{Gomez}},
  \bibinfo{author}{\bibfnamefont{J.}~\bibnamefont{Weinstein}},
  \bibinfo{author}{\bibfnamefont{K.}~\bibnamefont{Jones}}, \bibnamefont{and}
  \bibinfo{author}{\bibfnamefont{P.}~\bibnamefont{Lett}}, \bibinfo{journal}{New
  J Phys} \textbf{\bibinfo{volume}{8}}, \bibinfo{pages}{64}
  (\bibinfo{year}{2006}).

\bibitem[{\citenamefont{Salomon et~al.}(2014)\citenamefont{Salomon, Fouch\'e,
  Lepoutre, Aspect, and Bourdel}}]{PhysRevA.90.033405}
\bibinfo{author}{\bibfnamefont{G.}~\bibnamefont{Salomon}},
  \bibinfo{author}{\bibfnamefont{L.}~\bibnamefont{Fouch\'e}},
  \bibinfo{author}{\bibfnamefont{S.}~\bibnamefont{Lepoutre}},
  \bibinfo{author}{\bibfnamefont{A.}~\bibnamefont{Aspect}}, \bibnamefont{and}
  \bibinfo{author}{\bibfnamefont{T.}~\bibnamefont{Bourdel}},
  \bibinfo{journal}{Phys. Rev. A} \textbf{\bibinfo{volume}{90}},
  \bibinfo{pages}{033405} (\bibinfo{year}{2014}).

\bibitem[{\citenamefont{Burchianti et~al.}(2014)\citenamefont{Burchianti,
  Valtolina, Seman, Pace, De~Pas, Inguscio, Zaccanti, and
  Roati}}]{Burchianti2014}
\bibinfo{author}{\bibfnamefont{A.}~\bibnamefont{Burchianti}},
  \bibinfo{author}{\bibfnamefont{G.}~\bibnamefont{Valtolina}},
  \bibinfo{author}{\bibfnamefont{J.~A.} \bibnamefont{Seman}},
  \bibinfo{author}{\bibfnamefont{E.}~\bibnamefont{Pace}},
  \bibinfo{author}{\bibfnamefont{M.}~\bibnamefont{De~Pas}},
  \bibinfo{author}{\bibfnamefont{M.}~\bibnamefont{Inguscio}},
  \bibinfo{author}{\bibfnamefont{M.}~\bibnamefont{Zaccanti}}, \bibnamefont{and}
  \bibinfo{author}{\bibfnamefont{G.}~\bibnamefont{Roati}},
  \bibinfo{journal}{Phys Rev A} \textbf{\bibinfo{volume}{90}},
  \bibinfo{pages}{043408} (\bibinfo{year}{2014}).

\bibitem[{\citenamefont{Yamashita et~al.}(2017)\citenamefont{Yamashita,
  Hanasaki, Ando, Takahama, and Kinoshita}}]{PhysRevA.95.013609}
\bibinfo{author}{\bibfnamefont{K.}~\bibnamefont{Yamashita}},
  \bibinfo{author}{\bibfnamefont{K.}~\bibnamefont{Hanasaki}},
  \bibinfo{author}{\bibfnamefont{A.}~\bibnamefont{Ando}},
  \bibinfo{author}{\bibfnamefont{M.}~\bibnamefont{Takahama}}, \bibnamefont{and}
  \bibinfo{author}{\bibfnamefont{T.}~\bibnamefont{Kinoshita}},
  \bibinfo{journal}{Phys. Rev. A} \textbf{\bibinfo{volume}{95}},
  \bibinfo{pages}{013609} (\bibinfo{year}{2017}).

\end{thebibliography}

\end{document}